\documentstyle[aps,preprint,tighten,floats,epsf,rotate]{revtex}

\begin{document}
\draft

%
\title{Effect of pre-existing baryon inhomogeneities on the dynamics
  of quark-hadron transition.}
\author{
Soma Sanyal \footnote{e-mail: sanyal@iopb.res.in}}
%
\address{Institute of Physics, Sachivalaya Marg, Bhubaneswar 751005, 
India}
\vskip -0.5 in
%
\maketitle
\widetext
\parshape=1 0.75in 5.5in
\begin{abstract}

Baryon number inhomogeneities may be generated during the epoch when
the baryon asymmetry of the universe is produced, e.g. at the
electroweak phase transition. The regions with excess baryon number 
will have a lower temperature than the background temperature of the
universe. Also the value of the quark hadron transition temperature
$T_c$ will be different in these regions as
compared to the background region. Since a first-order quark hadron
transition is very  
susceptible to small changes in temperature, we investigate 
the effect of the presence of such baryonic lumps on the dynamics of
quark-hadron transition. We find that the phase transition is delayed 
in these lumps for significant overdensities. Consequently, we argue
that baryon concentration in these regions grows by the end of the 
transition. We briefly discuss some models which may give rise to 
such high overdensities at the onset of the quark-hadron transition. 

\end{abstract}
\vskip 0.125 in
\parshape=1 0.75in 5.5in

\narrowtext

\section{Introduction}

Phase transitions are supposed to have happened at different epochs in the 
early universe and one of the many possible consequences 
of these transitions is the 
generation of baryonic inhomogeneities. Most discussions in the
literature are about baryon inhomogeneities generated during the
quark-hadron transition \cite{witten,qcdbary,nucleate,ignatius}. One reason
for this is that baryon inhomogeneities naturally develop during 
the quark-hadron transition \cite{witten}. Another reason is that the
main importance of these inhomogeneities lies in the fact that if they
survive till the nucleosynthesis epoch, then they will affect the calculated
abundances of the light elements, leading to an inhomogeneous
big bang nucleosynthesis scenario. Since it is more probable that
inhomogeneities generated during the QCD transition would survive until
nucleosynthesis, that is why they are the ones which are mostly studied. 
However, baryon inhomogeneities can be produced at earlier stages as well.
For example, there are certain baryogenesis scenarios utilizing the 
electroweak transition which leave an inhomogeneous distribution of
baryons \cite{firorder}. These are scenarios where the baryogenesis 
occurs through strongly non-equilibrium processes. For example
ref.\cite{heckler} discusses 
the effect of baryon inhomogeneities generated by a first order
electroweak transition on the nucleosynthesis epoch. Apart from these
there are also defect-mediated baryogenesis models where the baryons  
generated are usually localized  
\cite{defemedia,ewstr,field,barriola,super}. These inhomogeneities are
expected to get homogenized by the effects of neutrino inflation
and baryon diffusion by the nucleosynthesis epoch\cite{branden}. So 
they usually do not affect the big-bang nucleosynthesis (BBN) calculations. 

However, even if these inhomogeneities do not survive until the stage 
of nucleosynthesis, they may still survive until the stage of 
quark-hadron phase transition. For example, in a detailed study by 
Jedamzik and Fuller\cite{jedamzik}, it was shown that overdensities
with large amplitude (of the order of ${n_b^\prime \over n_b} >
10^{3}$, where $n_b^\prime$ and $n_b$ are the baryon densities in the
overdense and underdense regions respectively) and lengthscales of the order of
$10^{-3}$cm (comoving at 100   
MeV), are dissipated very little (by the neutrino inflation process)
and may survive relatively undamped upto the stage when the
temperature of the universe is of order of 100 MeV, which is the scale
of the quark-hadron transition. 
At a later stage the mechanism for the damping of baryon
inhomogeneities becomes dominated by baryon diffusion which may
completely wipe out inhomogeneities which have a lengthscale less than 
the baryon diffusion length which is of the order of $10^{-1}$cm at
the nucleosynthesis epoch.
However we will argue in this paper that if the quark-hadron transition is of
first-order, and proceeds via bubble nucleation, then the 
baryon inhomogeneities which are already present during that time may
affect the dynamics of the phase transition. This is because
the bubble nucleation process in a first-order phase transition can
depend crucially on very small temperature changes.  In this work   
we study the effect of any pre-existing baryon inhomogeneities on the
dynamics of the quark-hadron transition. Since here the baryon 
inhomogeneity has to be present before the quark-hadron transition 
begins, we only consider the inhomogeneities generated at earlier 
stages, e.g., those generated during the electroweak epoch.

We consider the quark-hadron transition to be a first order
transition. The presence of the baryon inhomogeneities causes small
temperature fluctuations throughout the universe. 
The temperature of the inhomogeneities being less
than the surrounding temperature, neutrinos passing through them 
(for relatively small scale inhomogeneities) will tend to deposit 
heat in them. The lumps of inhomogeneities then inflate to achieve 
pressure equilibrium, thereby reducing the amplitude of the
inhomogeneity. At the same time, the inflation of the lump decreases
its temperature so the whole process is repeated again until the 
inhomogeneity is wiped out. This is the process of neutrino inflation, which
can efficiently erase baryon inhomogeneities depending on the 
amplitude and the scale of the inhomogeneity. As was shown by Jedamzik 
and Fuller \cite{jedamzik}, only for high overdensities the
inhomogeneities may survive for a large time. For example,
the inhomogeneities produced via certain models of baryogenesis at the 
electroweak scale may survive until the quark-hadron
transition.\cite{heckler}

 In the lumps with a higher baryon number density than the background, 
the value of the critical temperature for quark hadron transition
within the lump will be lower than that for the background \cite{mutc}. 
Now bubbles of hadronic phase only nucleate when there is sufficient 
supercooling.  We show that the combination of the effects of lower
critical temperature and the heat deposited by neutrinos is such that 
there is no nucleation of bubbles in the baryonic dense regions while
nucleation of bubbles gets completed in the surrounding region. 
So ultimately, we do not have a
homogeneous bubble nucleation scenario throughout the universe. 
Since at the site of the lumps the process of bubble nucleation is
delayed,  the bubbles nucleated outside the lump start expanding and 
reheat the universe to $T_c$ even before
nucleation has started in the baryon overdense regions. This region 
which already had an overdensity remains in the quark-gluon plasma (QGP)
phase. As the baryon number tends to remain in the QGP phase rather 
than in the hadronic phase \cite{witten}, this will lead to an 
increase in the overdensity in the already
existing lump. This will cause a further increase in pressure inside
the lump which will lead to the expansion of the size of the lump.
Thus even though the seed inhomogeneity was small in
size, it will give rise to a larger inhomogeneity as the phase
transition is completed. Basically the temperature fluctuation due to
the baryonic inhomogeneities prevents homogeneous nucleation of bubbles
throughout the universe. We show that due to this inhomogeneous nucleation of
bubbles, pre-existing baryon inhomogeneities grow in size as well as
in amplitude as the phase transition proceeds. (We mention here that it 
is likely that the bubble nucleation rate itself may be different in
the two regions, even with similar features of supercooling etc. If
bubble nucleation rate increases strongly with baryon density then one
cannot rule out a 
reverse situation where phase transition in the baryonic overdense
regions happen first. This will then cause a rapid decay of the
baryonic inhomogeneities, instead of it's growth. However in this
paper we ignore the dependance of bubble nucleation rate on baryon density.) 

In section II we present a detailed calculation of the characteristics
of the baryonic lumps, their temperature difference with the
background and their critical temperature. In section III, we briefly
review the time and temperature scales involved in a first order QCD
phase transition. Section IV then describes the effect of the baryonic 
inhomogeneities on the phase transition. In section V we briefly discuss 
some models in which such inhomogeneities may be generated before 
the quark hadron transition. Conclusions are presented in section VI. 

\section{Characteristics of the baryonic lumps.}

In the early universe, any small scale density inhomogeneity that is created  
achieves pressure equilibrium rapidly (typically with the speed of 
sound) with its surroundings. Before the QCD phase transition, the 
universe is in the QGP phase, with the pressure in a given region of space
due to quarks being,

\begin{equation}
P_q = {7 \over 4} N_q a T^4 + 9 N_q^{-1} {n_b^2 \over T^2}.
\end{equation}

\noindent where the second term on the right gives the contribution of
the baryon number to the pressure, $N_q$ is the number of
relativistic quark flavors at temperature $T$, and $a = {\pi^2 \over
30}$. A region of baryon inhomogeneity achieves pressure equilibrium
rapidly with its surroundings. 
The condition for pressure equilibrium between the outside and inside
the lump is given by,
\begin{equation}
{1\over 3} \epsilon + 9 N_q^{-1} {n_b^2 \over T^2} = {1\over 3}
\epsilon^\prime + 9 N_q^{-1} {n_b^{\prime 2} \over T^{\prime 2}}   
\end{equation}
Here ${\epsilon \over 3}$ and ${\epsilon^\prime \over 3}$ are the
radiation pressures due to all the relativistic particles including leptons and
photons in the two regions (with $ \epsilon = g_{eff} a T^{4}$, and
$g_{eff}(\simeq 51)$ is the effective degrees of freedom in the
quark-gluon plasma phase). For the relevant values of baryon  
number, one can see from Eq.(2) that among the two regions whichever has 
a higher baryon number ($n_b^\prime$) must have a lower temperature.
Replacing $T^\prime = T + \delta T$ in Eq.(2)  one can calculate  
the temperature difference ${\delta T \over T}$  between the baryon 
over-dense and under-dense regions \cite{jedamzik}. 
It is given by,

\begin{equation}
{\delta T \over T} = - {27 \over 4 g_{eff} a T^4} {n_b^2 \over T^2} {1
  \over N_q} [({n_b' \over n_b})^2 - 1]
\end{equation}
 
 For $T \sim 170$ MeV, and for $({n_b' \over n_b})^2 >> 1$ we get,

\begin{equation}
\Delta T^{\prime} \equiv {\delta T \over T} = - 3 \times
10^{-19}\times ({n_b' \over n_b})^2.
\end{equation}

 This equation gives the dependence of the temperature difference of the
overdense region from the background on the magnitude of the
overdensity in the region. We will see that for  sufficiently large
values of $({n_b' \over n_b})^2$ the difference in temperature between 
the inside of the lump and its surroundings can be significant enough
so that heat deposition in these regions by neutrinos can 
disrupt the usual dynamics of a first order quark-hadron phase
transition. This is because 
the dynamics of a first order phase transition depends crucially on
temperature differences of even very small scales. As we will see below,
this happens if we have  $({n_b' \over n_b}) > 10^{7} $.

Even though the temperature in the baryonic overdense lump is lower
than the background temperature, this temperature difference is 
relatively small (as we will see below). Thus, with higher baryon density
inside the lump, the baryon chemical potential $\mu$ inside the lump is also 
larger than the corresponding value in the background region.
As the critical temperature for the quark-hadron phase
transition depends on the chemical potential,  the difference in the
chemical potentials will also cause a difference in the values of the 
critical temperature, with $T_c^\prime$ and $T_c$ denoting the values
of critical temperature for the overdense and the background regions
respectively.  
Unless a region supercools by a certain amount below the critical 
temperature (suitable for the region under consideration), bubble 
nucleation will not start. As the critical temperature is different in 
the two regions, bubble nucleation may also start in the two regions at 
two different times, unless the temperature difference between the
two regions exactly compensates for the effect of difference between
$T_c^\prime$ and $T_c$. We will express the value of the chemical 
potentials in a given region in terms of $\eta = {n_b \over s}$ where
$n_b$ is the baryon number density and $s = {2 \pi^2 \over 45} g_{eff}
T^3 $ is the entropy density. $T$ is the temperature of the region.  
 We get \cite{eta}, 

\begin{equation}
{\mu \over T} \sim  12 \times \eta,
\end{equation}

 Using Gibb's criterion for a first order phase transition, we equate
the pressure in the quark-gluon phase to the pressure in the hadronic 
phase to determine the corresponding critical temperature\cite{cleyman}.
With non-zero chemical potential we get \cite{cleyman},

\begin{equation}
P_q = P_{\pi} + P_N
\end{equation} 

where

\begin{equation}
P_q = {37 \over 90} \pi^2 T^4 + ({\mu \over 3})^2 T^2 + {1 \over 2
  \pi^2} ({\mu \over 3})^4 - B, 
\end{equation} 

\begin{equation}
P_{\pi} = {3 m^2 T^2 \over 2 \pi^2} \sum_{k=1}^ {\infty} {K_2({km
    \over T}) \over k^2},
\end{equation} 
and
 
\begin{equation}
P_N = {2 M^4 \over 3 \pi^2} \int_0^1 {u^4 du \over (1-u^2)^3}
[f(u;T,\mu) + f(u;T,-\mu)].
\end{equation}

Here $P_q$ is the pressure in the quark-gluon phase, while the right
hand side of Eq.(6) gives the total pressure in the hadronic
phase (taken as the pressure of pions and nucleons). Note that here
 we consider only the QCD degrees of freedom, as contribution from
other particle species cancels out. 
$K_2({km\over T})$ is the modified Bessel function of the second kind   
and the function $f(u;T,\mu) = {({exp{[M/(1-u^2)^{1/2} - \mu] \over
T}}+1)}^{-1}$. $M = 940$ MeV is the nucleon mass and $m =  140 $ MeV is 
the pion mass. Solving Eq.(6) we can get the value of $T_c$ for a given
value of $\mu$.  

 We mention here that we use Eqs.(6)-(9) only to get an order of magnitude 
estimate of the shift in the transition temperature as a function of
$\mu$. This is primarily determined by $\mu$ dependant terms in
Eqs.(6)-(9). As lattice results are not available for the range of
values of $\mu$ relevant to us which could constrain these terms,
we use Eqs.(6)-(9) for our order of magnitude estimates. For $\mu = 0$, 
lattice results indicate small values of surface tension. With that, 
supercooling will be even smaller than used in our paper. As we
have discussed below, this does not affect our conclusions (as long as 
the transition still remains first order), as with smaller supercooling,
our results should apply even for baryon inhomogeneities
of smaller magnitude.

\section{First order QCD phase transition.}

We now briefly describe the time and temperature scales involved in 
a first order QCD phase transition in the early universe as has been
discussed extensively in the literature. A first order phase transition
proceeds with the formation of bubbles of hadronic phase within the
QGP phase. These bubbles then expand and gradually convert all the QGP 
phase to the hadronic phase. Since the critical size of the bubbles is
too large near $T_c$, the universe has to supercool slightly below the
critical temperature for the rate of  nucleation of bubbles to be adequate.
Essentially, the rate of bubble nucleation should become significant
compared to the expansion rate of the universe. The amount of 
supercooling required and its duration depends on quantities such as 
the latent heat  of the transition $L$, the surface tension $\sigma$
of the interface etc. Using results from lattice
calculations,\cite{nucleate,ignatius,lattice} ($L \sim 4T_c^4$ and
$\sigma \sim  0.015 T_c^3$) we can estimate that the amount of
supercooling $\Delta T_{sc} \equiv {T_{c}-T{sc} \over 
T_c}$ required for the nucleation of bubbles is \cite{nucleate} of the
order of $\sim 10^{-4}$. We mention here that it has been argued in
the literature that the amount of supercooling may be smaller by many
orders of magnitude \cite{gavai}. As we will see later, for smaller
supercooling our results imply that the quark-hadron transition can be 
affected by baryon inhomogeneities of even much smaller magnitudes. 
After nucleation, the bubbles begin to 
expand releasing latent heat which reheats the universe
back to the critical temperature. This happens in a very short time
and temperature interval
and bubble nucleation is shut off after that. The time scale for this
is given by $\delta t \sim 10^{-5} t_H$ (where $t_H \sim 10^{-6}$ sec 
is the Hubble time at QCD transition) and the temperature interval is
given by $\Delta T_n \sim 10^{-6}$ (see also
ref.\cite{nucleate,ignatius}). No nucleation  
of bubbles can happen after this and
the transition proceeds by the very slow expansion of the already nucleated  
bubbles as the universe expands and cools, gradually transforming the QGP
phase to the hadronic 
phase. This phase of slow expansion is usually referred to as the
``slow combustion 
phase''\cite{witten}. It is this phase which is very different in
models where there is an inhomogeneous nucleation of
bubbles. Evidently some parts of the universe enter the slow
combustion phase before other parts. If a large portion of the
universe is in the slow combustion phase, then the large amount of
latent heat generated by the expansion of bubbles prevent nucleation
of bubbles in the other parts (which have not entered the slow
combustion phase as yet). This modifies the nature of the phase
transition drastically.

We again emphasize that it is likely that the rate of bubble
 nucleation in the  
baryon over-dense region will itself be different from the rate of bubble 
nucleation in the outside regions even with similar factor of
supercooling etc. This in principle could even reverse the sequence of
transitions in the two regions. However, for simplicity we ignore this
possibility in the present work, and only focus on the temperature
differences between the two regions and resulting differences in the
onset of bubble nucleation. We also assume that the supercooling
required for starting off the phase transition in the QGP phase is of
the same order of magnitude for different chemical potentials in our
case. A more through investigation will have to include both these
considerations.     

\section{Effect of the inhomogeneities on the phase transition.}

Now we discuss in detail how exactly the phase transition is affected
by the presence of the baryon inhomogeneities. We consider the stage when the
background region surrounding the inhomogeneities reaches sufficient
supercooling  
for bubble nucleation to start. For bubble nucleation to start in a
baryon dense region, it should also achieve sufficient supercooling. 
We consider the situation when the baryon inhomogeneities are
at least of order, ${n_b' \over n_b} = 10^7 $.  This is because only for
such magnitudes of baryon inhomogeneities, we find a significant effect
on the dynamics of quark-hadron transition. (Note that this is for
$\Delta T_{sc} \sim 10^{-4} T_c$. For smaller values of $\Delta
T_{sc}$, as in ref.\cite{gavai}, similar effects will be found for
even smaller values of ${n_b^\prime \over n_b}$.) We will later briefly discuss
how such inhomogeneities could possibly arise. With $T \sim 170 $MeV,
the value of $\mu$  (using Eq.(5)), for ${n_b' \over n_b} = 10^7 $ comes out 
to be $\mu \sim 14 $ MeV. Compared to this, the value of
$\mu$ outside is $\sim 10^{-6} $ MeV. We have taken the background
value of $\eta \sim 7 \times 10^{-10}$. The value of critical temperature
$T_c$ at zero chemical potential is $T_c \simeq 172$ MeV (From Eqs.(6)-(9)). 

Using the values of $\mu$ corresponding to the values of $n_b$ and
$n_b^\prime$,
we can calculate the corresponding values of the critical temperatures
using Eq.(6). For $\mu = 10^{-6} $ MeV corresponding to the background, 
the fractional change in the value of critical
temperature (compared to the zero chemical potential case) is
$\Delta T_c/T_c \ll 10^{-5}$. As we have mentioned before, values of
fractional  
temperature differences which are relevent for the phase transition
are of order $10^{-6} - 10^{-4} $  Therefore, the change in the
critical temperature for 
the background chemical potential is negligible and the process
of phase transition in the regions outside the baryonic lumps remains
unaffected. As the background regions supercool sufficiently to the
required temperature $T_{sc}$ (with ${T_c-T_{sc} \over T_c} \simeq 
10^{-4}$), the region inside the baryonic lump also cools by
approximately same factor. At any stage, the difference between the 
background temperature $T$ and the temperature $T^\prime$ inside the 
lump is given by Eq.(4). With the values of the chemical potential
given above, and for temperatures close to $T_c$, we get,

\begin{equation}
{T^\prime - T \over T} \equiv \Delta T^\prime  \simeq - 4 \times 10^{-5}.
\end{equation}
 
At the same time the critical temperature for the phase transition to occur in
such a lump is given by $T_c^\prime$ where, using Eq.(6), 

\begin{equation}
{T_c^\prime - T_c \over T_c} \equiv \Delta T_c^\prime \simeq - 4
\times 10^{-5}
\end{equation}

Thus the temperature difference between the lump and the background is 
essentially the same as the difference in the values of $T_c$. 

Assuming similar supercooling factor of $10^{-4}$ for the baryonic lump, bubble
nucleation there cannot start until the temperature drops to a value
$T_{sc}^\prime \simeq (1 - 4 \times 10^{-5} - 10^{-4})\times
T_{c}$. Here, the factor of $4 \times 10^{-5}$ comes from Eq.(11), 
while the other factor of 10$^{-4}$ arises from the requirement of
sufficient supercooling for bubble nucleation. 

Using the relationship between $T_c$ and $T_{sc}$ we get,
\begin{equation}
T_{sc}^\prime =  (1 - 4 \times 10^{-5}) \times T_{sc}
\end{equation}

Thus the difference between $T_{sc}^\prime$ 
and $T_{sc}$ is of the same order as the temperature difference
between the lump and the background given by Eq.(10). 
For simplicity we assume that nucleation rates are not significantly
affected by the relatively large value of $\mu$ inside the lump. Thus,
one will conclude that when background temperature reaches the value
$T_{sc}$, the temperature in the lump may also be close to the
corresponding supercooling temperature $T_{sc}^\prime$ 
for the lump. We mention here again
that it is entirely possible that nucleation rates change
significantly \cite{kpst} as a function of $\mu$, which can strongly
affect our discussion. We hope to discuss this issue in a future work.

  The discussion above neglects one important factor, whose effect is
to delay the phase transition process in the baryonic lumps as 
explained below. Since the dense region has a lower temperature
than the outside, neutrinos will be constantly pumping in heat which
will keep on raising the temperature of the lump.
This will temporarily increase the temperature of the lump
until the pressure equilibrium relaxes the lump, thereby decreasing 
its temperature again in accordance with Eq.(3).   
However, the lump will require a certain amount of time to 
regain its pressure equilibrium. When the outside 
temperature reaches $T_{sc}$ then even though we are allowing for 
the possibility that the lump also could reach the corresponding 
supercooling temperature $T_{sc}^\prime$, 
it will be possible only when the lump maintains pressure equilibrium
with the surroundings. Until pressure equilibrium is achieved, the
temperature of the lump will rise above the value $T_{sc}^\prime$.
This implies that during a period until pressure equilibrium
is re-established, no nucleation of bubbles will be possible 
in the overdense region. 

 The timescale for the lump to attain pressure equilibrium will be of
order $\Delta t_p = R/c_s$, where $c_s$ is the sound velocity and $R$ 
is the size of the lump. If we take the size of the lumps $R \sim 1$ cm,
then with the sound speed  $c_s = {1 \over \sqrt{3}}$,
the time for attaining pressure equilibrium will be $\Delta t_p \sim 6 
\times 10^{-11}$ sec. This is  larger than the time taken to
complete the bubble nucleation process in the outside region ($\Delta
t_{n} \sim 10^{-5} t_H = 10^{-11}$ sec). Since the temperature of the
lump will be temporarily increased during this time period, the bubble
nucleation process may not start inside the lump, depending on the magnitude
of this temperature increase. 
Note that as we are interested in the temperatures very close to
the critical temperature, the relevant sound speed will typically be
much smaller than the value ${1 \over \sqrt{3}}$. The duration $\Delta t_p$
for which the bubble nucleation will be delayed inside the lump should
therefore be much longer than the value given above. In fact as shown
in ref.\cite{kampfer}, the sound speed near $T_c$ can become very
small (e.g. even smaller than $0.2$),
whereby the time for attaining pressure equilibrium will become at least 
$10^{-10}$ sec, which is much larger than the time taken to complete
the bubble nucleation process in the region outside. Also note that
heat deposited in the  
lump is carried by relativistic neutrinoes so the timescale for
neutrinoes to pass through the lump always remains much shorter than
the time scales discussed here \cite{jedamzik}.     

 We now obtain the temperature rise due to the heat deposited in the 
lump for a time duration $\Delta t_n$ which, as we have discussed
above, is  the timescale of bubble nucleation in
the surrounding region ($\Delta t_n = 10^{-5} t_H$).

The heat deposited  by the neutrinos in a given volume depends on
the size of the lump $R$ compared to the neutrino mean free path $\lambda$ 
at that temperature. Since the neutrino mean free path around the
quark-hadron transition is few cms, hence in this case $R \sim
\lambda$. For $R \sim \lambda$, the neutrino radiation can be
approximately considered to be perfectly absorbed throughout the
lump\cite{hogan}. Hence we have, 

\begin{equation}
{dE \over dt } \simeq 4 \pi R^2 \Phi
\end{equation}
where $\Phi$ is the net energy flux into the lump.

\begin{equation}
\Phi = \sum_i \rho_i(T) {\delta T \over T}, 
\end{equation}
$\rho_i(T)$ being the energy density of the neutrinos, 
($i$ representing the type of
neutrinos)  and ${\delta T \over T}$ is given by equation (4).
Keeping the volume ($V = {4\over 3} \pi R^3$) constant, the
temperature rise is given by, 

\begin{equation}
V g_{eff} a 4 T^3 {dT \over dt} =   4 \pi R^2 g_{\nu} a T^4 \times 3 \times
10^{-19} ({n_b' \over n_b})^2 
\end{equation}

Here $g_{\nu}$ is the effective degrees of freedom for the
neutrinos and is taken to be $ 6 \times {7 \over 8}$. On the right
hand side of equation (15), we have 
contribution only from the neutrinos because they are the only
particles depositing heat in the lumps. However this heat is absorbed
by all the particle species in the lump, hence on the left hand side
of Eq.(15) we have considered all the particle species.  
Substituting all the values we get,

\begin{equation}
{dT \over T} = 0.4 \times 10^{-19} ({n_b' \over n_b})^2 \times {dt
  \over R} 
\end{equation}

For $R = 1 $ cm and $dt = \Delta t_n = 10^{-11}$ sec we get,

\begin{equation}
 {dT \over T} \sim   10^{-20} ({n_b' \over n_b})^2 \\
                 ~~ > 10^{-6}  ~~~{\rm for}  {n_b' \over n_b} > 10^7 
\end{equation}

 We note that the temperature rise due to heat deposited in a given 
baryonic lump during the time interval $\Delta t_n$ is of the same 
order of magnitude as $\Delta T_n$ which is the temperature interval below
$T_{sc}$ in which bubble nucleation process shuts off. This implies
that while outside region reaches it's lowest temperature $(\Delta
T_{sc} -\Delta T_{n})$ before it starts reheating due to latent heat
release , the temperature inside the lump can barely reach down to
$T_{sc}^\prime$.
We, therefore, conclude that inside a lump of size $R \simeq 1$ cm, and an 
overdensity corresponding to ${n_b'\over s}$ of the order of $10^{-3}$,
it is not possible to have any bubble nucleation when the
surrounding region undergoes bubble nucleation. The region in the lump 
cannot reach its respective nucleation temperature while nucleation
starts and completely shuts off outside. Expanding bubbles in the
outside region will release latent heat, thereby raising the temperature of
the outside region back to the critical temperature $T_c$. Importantly, 
heat transport by neutrinos will also further raise the temperature 
inside the lumps, implying that bubble nucleation will remain
shut-off inside the lump  while the outside region undergoes the
slow combustion phase \cite{witten}. It is important to mention here
that if smaller values of sound velociy are taken into account then
even with smaller size lumps  these conditions hold. With a smaller size 
lump (say $ \sim 0.1$ cm) and smaller velocity of sound ($ \sim 0.1$),
$\Delta t_p $ remains of the same 
order, so that the region will not be able to achieve pressure
equilibrium in the interval $\Delta t_{n}$. But for a smaller size
lump ($R < \lambda$), Eq.(16) 
will be modified by a factor of $ R \over \lambda $ on the right hand
side as only this much fraction of neutrino energy will get deposited
in the lump. However this does not change the value of $dT$ in Eq.(17)
as can be seen 
by multiplying the R.H.S. of Eq.(16) by ${R\over \lambda}$. It will depend
on $\lambda$ instead, and for temperatures around 100 MeV we have
$\lambda \simeq 1$ cm.  So for lumps with $ R < \lambda $, we can
reach the same conclusion as before. 

There only remains the case of $ R > \lambda $. For this case neutrino 
heat conduction is small \cite{hogan}, but the size of the lump being
quite large, $\Delta t_p $ will be much greater than $\Delta t_n$. So
the lump will be able to come to pressure equilibrium in a time which
will be much larger than the
time in which nucleation of bubbles is complete in the outside
region. So we may conclude
that essentially for any size lump there is no nucleation of bubbles
possible within the overdense region. 

Baryonic lumps having ${n_b'\over s}$ of the order of $10^{-3}$ is the 
lower limit for lumps which would affect the QCD phase transition
(again , for $\Delta T_{sc} \sim 10^{-4}$). 
If we consider lumps with higher overdensities, ${\delta T \over T}$
in Eq.(3)
will be larger. At the same time the critical temperature will also 
go down. Important thing is that due to larger $\delta T$, the heat
deposition by the neutrinoes will increase (see Eq.(14)). However 
for the background region the amount of supercooling required and the
temperature 
interval below $T_{sc}$ for bubble nucleation to shut off (i.e $\Delta 
T_{sc}$ and $\Delta T_{n}$) will remain
the same. Therefore due to larger value of dT in Eq.(17), 
it will be even more difficult for bubble nucleation to take place
within these lumps.  
For example, if we have ${n_b'\over n_b} = 10^8 $, then ${\delta T
  \over T} = - 4. \times 10^{-3}$. But the main factor responsible for
suppressing bubble nucleation in the overdense region is the heat
deposited by the neutrinoes. This will now be much
greater than the temperature interval of bubble nucleation
($\Delta T_{n}$). In this case the temperature increase due to the
heat deposited by the neutrinoes ($dT \over T$)
is of the order of $ 10^{-4}$ which is greater than $\Delta T_{n}$ by
two orders of magnitude. Hence there is no way that bubble nucleation
can start in these overdense lumps.     So if there are
inhomogeneities present which have  ${n_b'\over s} \ge 10^{-3} $,
there is every possibility that the whole phase transition will
proceed by the inhomogeneous nucleation of bubbles in the various
regions of different baryon densities. So the lumps tend
to remain in the QGP phase while the rest of the universe undergoes
the phase transition to the hadronic phase.

 As first discussed in ref.\cite{witten} (see, also, ref.
\cite{qcdbary,brn}), baryon number tends to remain in 
the quark-gluon phase rather than the hadronic phase where they are
carried by the more massive hadrons. Since the baryon overdense region 
will be in the QGP phase while the outside region hadronizes, baryon number
will tend to concentrate in this region. Thus, in our model, the 
inhomogeneity in the lump is not depleted as long as the quark-hadron
transition continues because the baryon number keeps getting 
concentrated inside the remaining QGP regions which are the regions
of baryonic lumps. The total amount of baryons which get concentrated
inside such lumps will depend on the detailed geometry of bubble
collisions and coalescence. For example, even in the regions outside
the lumps, spherical QGP regions will form with increased baryon
concentration due to QGP regions getting trapped in between coalescing 
bubbles.\cite{witten} It has been discussed in the literature that the largest
separation \cite{ignatius,heckler} between these inhomogeneities 
which are formed during the quark - hadron transition is of the order
of few centimetres for homogeneous nucleation of bubbles and of the order of a
metre if inhomogeneous nucleosynthesis (due to temperature  
fluctuations) is considered. \cite{ignatius} One thing that we have to 
consider is the fact that in our model, the pre-existing baryon
inhomogeneities have  
a very high ${n_{b} \over s}$. Hence the fraction of volume occupied by these
inhomogeneities must be very small, otherwise most of the baryon
number would anyway be 
concentrated in these regions. But again if the number of these
pre-existing inhomogeneities be very few and far between then during
the transition they would focuss very little of the total baryon
number. To check whether it is okay to have such high
inhomogeneities and also focuss a fairly large percentage of baryon
number due to the previously discussed phenomenon, 
we make an estimate of the distance of separation $l$ and radius $R$
of the pre-existing inhomogeneities such that only a small fraction of 
the baryon number in the universe before the quark-hadron transition
is concentrated in these regions. 

The fraction of volume occupied by the high density regions is roughly
given by $f_{H} = ({R \over l})^3$. We may therefore write,

\begin{equation}
f_{H} n_{H} + (1 - f_{H}) n_{L} = n_{b}
\end{equation}

Here, $n_{H}$ and $n_{L}$ are the baryon densities in the high density
regions and low density regions respectively, while $n_{b}$ gives the
total average baryon density. For $R \ll L$, we will have $f_{H} \ll 1
$, hence, 
$n_{L} \simeq n_{b} - f_{H} n_{H}$. Since we have $n_{b} = 10^{-10}
\times s $ and $n_{H} = 10^{-3}\times s $, if we want to have $n_{L} \sim
10^{-10}\times s $  we must have, $f_H < 10^{-7}$.
So the condition for the pre-existing inhomogeneities to have a
negligible effect on the total baryon number turns out to be,
\begin{equation}
({R \over l})^3 \le 10^{-7}.
\end{equation}
As we have discussed before, the size of the inhomogeneity can be
taken to be as small as 0.1cm, thus we get $l \ge 20 $ cms. So it is
possible to have pre-existing inhomogeneities with radius 0.1 cm and
separated by a distance scale of 20 cm at the onset of the
quark-hadron transition. 
Since there will be two processes going on simultaneously, focussing
of baryon number in pre-existing inhomogeneities and generation of new
regions of baryon overdense regions in the inter-bubble spacings, we
compare the lengthscales involved in the two cases.  In homogeneous bubble 
nucleation it is known that \cite{ignatius,heckler} the separation of baryon
over-dense regions formed as the 
bubbles coalesce is few cms. This is not too small compared to the
length scale $l \sim 20$ cm of pre-existing baryonic lumps. Further,
with an even smaller velocity of sound (near $T_c$), we can even have
smaller values of R and smaller $l$. For example with $R \sim 0.01$
cm, we will get $l  
\sim 2 $ cm. This is the same as the expected separation between the
QGP droplets generated during the phase transition. In such situations 
at least 50\% of the baryon number can get concentrated in these
pre-existing inhomogeneities. Also note that the actual fraction of
baryon concentration in the pre-existing inhomogeneities will depend
on detailed geometry of bubble collision in the background region. It
is possible that for larger values of $l$ also, a good fraction of
baryons may get concentrated in these pre-existing lumps.

One more important thing to keep in mind is that the proton diffusion
length scale (which is the dominant mechanism for erasing out
inhomogeneities after neutrino decoupling) depends crucially on the
amplitude of the overdensities. For amplitudes greater than 100 the
proton diffusion length scale decreases rapidly,\cite{heckler} going
down to even 0.001 cm for amplitudes of the order of $10^{6}$. So even 
though the size of our pre-existing inhomogeneities will be much
smaller compared to the the length of the newly generated
inhomogeneities ($ \sim 2 $cms) they will not be erased by proton
diffusion because of their high amplitudes.

\section{Generation of inhomogeneities before the QCD transition.}

 All this will only be possible if there are such large 
overdensities present just before the QCD phase transition. We now
discuss some processes which may generate such overdensities before
the QCD transition. However we mention here that the principal aim of
our work was to show how large pre-existing baryon inhomogeneities may 
affect the dynamics of a first order quark-hadron transition. Whether
such high overdensities exist or not is still not clear. As we have
mentioned earlier, 
the inhomogeneities generated at an early stage may or may not survive until
nucleosynthesis, but as shown in ref.\cite{jedamzik}, 
sufficiently high overdensities may remain more-or-less unchanged between the
electroweak epoch and the quark-hadron transition. Most of these
inhomogeneities usually do not survive upto the nucleosynthesis epoch
if their lengthscale  at any stage is less than the proton diffusion
length. \cite{heckler,jedamzik}. If such
inhomogeneites are present then our work shows that they will
definitely affect the dynamics of the QCD phase transition. We now
suggest some possibilities as to how such inhomogeneities may be
generated.

One of the processes through which one may be able generate such
inhomogeneities is by 
electroweak strings. Electroweak strings, which may be formed during
the electroweak phase transition, are unstable and are expected to
decay rapidly.  
There is a lot of literature which discusses the generation of
baryon numbers by electroweak strings\cite{ewstr,field,barriola}. The changing
helicity of an electroweak string network can generate a baryon 
to entropy ratio of order $10^{-4} \epsilon$, where $\epsilon$ 
is the CP violating factor \cite{field}. Though this asymmetry is
generated when the network consists of  
a large number of loops and gives the average over the horizon volume,
there may be a possibility that over certain regions the baryon number
concentration may be larger than at other places. This possibility will 
definitely arise when the electroweak strings are metastable and 
manage to survive to lower temperatures. Recently several authors
have indeed discussed such a possibility. \cite{stability}. If these
strings are stable upto around 1 GeV, then strings, which decay late, will
typically be of large size and will
generate overdensities of larger sizes, well separated and far from each
other. However since the number of such large strings  will be very
small at later stages, the baryon number they generate will not affect 
the overall baryon-to-photon ratio. (The bulk of baryons will still
be generated by small strings decaying right after the electroweak 
transition.) The larger strings decaying late will generate  sharp
peaks of overdensities in localized regions. 

Another way in which electroweak strings may generate baryons is by
their decay, especially in a magnetic field \cite{barriola}. Large
overdensities may be generated by these strings over small length
scales ($\sim 10^{-10}$cms), in the presence of a magnetic field if
they decay at the electroweak epoch itself. These densities will
generally diffuse out to some extent but as shown in \cite{jedamzik}
they may  not be wiped out completely. Again if such strings
decay later in the presence of strong magnetic fields they may generate
larger inhomogeneities locally. As we are interested only in local baryon
inhomogeneities the volume suppression factor will not be of
concern here.
   
There is also the possibility of large superconducting string loops
(formed around the TeV scale)
decaying into vortons and generating baryon number when these vortons
subsequently decay. In these
models the baryon asymmetry generated earlier is protected from
sphaleron wash-out and released later when sphaleron processes fall
out of equilibrium\cite{super}(i.e after the electroweak
transition). Apart from strings, other defects 
like the domain walls also generate baryon number\cite{dw}. Black holes
evaporating between the 
EW epoch and the QCD phase transition are another source of generation 
of such baryon inhomogeneities\cite{bh}. Even a first order
electroweak phase transition seeded by impurities can generate
inhomogeneities of lengthscale $10^{-3}$ cm at the electroweak epoch,
(which translates to about $1$ cm at the QCD epoch)  
\cite{heckler} but the amplitude of these inhomogeneities are only as
high as $10^{4}$. Altogether we see that there
are several possibilities in which large baryon inhomogeneities may be
generated before the quark-hadron transition and we will postpone the
details of the generation of such large inhomogeneities for a later
work.

\section{Conclusion}

In conclusion, we have studied the effect of pre-existing baryon
inhomogeneities on the dynamics of a first order quark-hadron
transition. Our studies show that though the temperature in the 
lumps of high baryon densities is lower than the outside temperature,
the bubble nucleation temperature at these high density lumps  is also 
lower, and due to heat deposition by neutrinoes, the process of 
phase transition is delayed in these regions. To demonstrate this we
have estimated  the 
difference in the critical temperature between the outside region and
the baryon rich region as also the temperature rise due to neutrino
heat deposition. We show that the
baryonic regions do not achieve enough supercooling for bubble
nucleations to effectively start in these regions. (We mention here again
that if smaller supercooling is required for
nucleation of bubbles, e.g. as discussed in ref. \cite{gavai}, 
our scenario remains unchanged. In
fact if supercooling required is much less than $10^{-4}$, our mechanism
will also work for overdensities much smaller than $10^{7}$.)

 These regions do not reach sufficient supercooling due to 
heat  being continuously deposited in the lumps by neutrinos and
the time scale for the lumps to inflate  
and achieve pressure equilibrium is at least of the
same order of magnitude as the time required for bubble nucleation to
be completely shut off in the outside region. Once bubble nucleation
is completely shut off and the outside region has reached the slow
combustion phase the latent heat released by the expanding bubbles
prevents further nucleation of bubbles everywhere, including the 
baryon dense regions. The final result is that the bubble nucleation
never takes place in the regions of baryonic lumps. 
Since baryon number tends to stay in the quark phase, the
baryon number in the already overdense region increases as the region
outside gets converted to the hadronic phase, and since the
size of the inhomogeneity also increases due to neutrino inflation,
we may get large baryon overdense regions at the end of the
quark hadron transition.  Thus we see that a smaller baryon
inhomogeneity may lead ultimately to a much larger baryon
inhomogeneity. This is very different from the conventional picture of
the evolution of baryon inhomogeneities in the early universe where the
region of inhomogeneity will anyway  increase in size while the 
amplitude of the  inhomogeneity decreases \cite{jedamzik}. 
In our work we have shown that during the quark-hadron
transition the inhomogeneity will still increase in size but its
amplitude will not decrease, on the other hand its amplitude may
increase substantially by the end of the transition. In our
calculation we have neglected the difference in nucleation rates
because of the difference in chemical potential in the baryon
overdense regions. For a more complete
study both the difference in nucleation rate and the amount of
supercooling in the two regions need to be taken into account. If the  
nucleation rate depends strongly on the chemical potential, (say, it
increases with chemical potential)  then 
as we have mentioned earlier, it may so happen that phase transition 
occurs in the inhomogeneities first and then in the rest of the
region; in which case the inhomogeneities will not grow but will
dissipate away. Recently there has been some lattice calculations of
the QCD phase transition at finite $\mu$ and finite T. \cite{fodor}
Though for small values of $\mu$ and finite T the transition is
expected to be a crossover, however as $\mu$ increases there happens
to be a critical $\mu$ beyond which it becomes a first order
transition. The endpoint is still at too large a value of $\mu$ but it 
is expected to move closer to the $\mu = 0 $ value for more realistic
quark masses. Also we mention that it is possible that the value of
$\mu$ in the baryon inhomogeneities may be greater for large overdensities.

We have also briefly mentioned various possibilities about 
the generation of such high
overdensities. There are many ways by which localized regions of high
densities may be generated between the electroweak epoch and the quark 
hadron transition. We have not given any specific estimates but
have commented 
on a variety of models dealing mostly with topological and
non-topological defects. In a later work we hope to address this issue 
in more detail.

\vskip .2in
\centerline {\bf ACKNOWLEDGEMENTS}
\vskip .1in

  I am very thankful to Ajit M. Srivastava for his useful comments
  and detailed discussions regarding every aspect of the paper. I am 
  also thankful to Rajarshi Ray and Biswanath Layek for many
  interesting discussions and helpful suggestions. I am also thankful
  to Sanatan Digal for his useful comments. 



\begin{thebibliography}{99}

\bibitem{witten} E. Witten, Phys. Rev. {\bf D30}, 272 (1984).

\bibitem{qcdbary} J.H. Applegate and C.J. Hogan, Phys. Rev. {\bf D31},
3037 (1985); J.H. Applegate, C.J. Hogan, and R.J. Scherrer,
Phys. Rev. {\bf D35}, 1151 (1987); G.M. Fuller, G.J. Mathews, and C.R. Alcock,
Phys. Rev. {\bf D37}, 1380 (1988); K. Kajantie and H. Kurki-Suonio, 
Phys. Rev. {\bf D34}, 1719 (1986); J. Ignatius, K. Kajantie,
H. Kurki-Suonio, and M. Laine, Phys. Rev. {\bf D50}, 3738
(1994). 

\bibitem{nucleate} K. Kajantie, Phys. Lett. {\bf B285}, 331 (1992);   
B. Layek, S. Sanyal and A.M. Srivastava, Phys. Rev.{\bf D63}, 83512, (2001).

\bibitem{ignatius} J. Ignatius and D.J. Schwarz, Phys. Rev.{\bf L86},
  2216, (2001)  
 
\bibitem{firorder} A. Cohen, D. Kaplan and A. Nelson, Phys. Lett. {\bf
B245}, 561, (1990);  Nucl.Phys.{\bf B349}, 727, (1991); Phys. Lett. {\bf
B263}, 86, (1991); L.D. McLerran, M.E. Shaposhnikov, N. Turok and
M.B. Voloshin, Phys. Lett. {\bf B256}, 451, (1991);
L.D. McLerran,Phys. Rev. {\bf L62}, 1075, (1989); M.E. Shaposhnikov,
JETP Lett. {\bf 44}, 465, (1986) ; Nucl. Phys. {\bf B287},
757, (1987); Nucl. Phys. {\bf B299}, 797, (1988); N. Turok and P. Zadrozny, 
Phys. Rev. {\bf L65}, 2331, (1990); Nucl. Phys. {\bf B358}, 471,
(1991).

\bibitem{heckler} A. F. Heckler, Phys. Rev. {\bf D51}, 405 (1995).



\bibitem{defemedia} A.C. Davis and R. Brandenberger, Phys. Lett. {\bf
   B308},79 (1993), M. Joyce and T. Prokopec, Phys. Rev. {\bf D57}, 6022 
 (1998); R. Brandenberger, A.C. Davis and M. Trodden, Phys. Lett. {\bf B355},
123 (1994);  R. Brandenberger, A.C. Davis, T. Prokopec and M. Trodden
 Phys. Rev. {\bf D53}, 4257 (1996). 


\bibitem{ewstr} J. Dziarmaga, Phys. Rev. {\bf D52}, R569 (1995);
M. Nagasawa, Astropart. Phys {\bf 5}, 231 (1996); 
M.Sato, Phys. Lett. {\bf B376}, 41 (1996);  M. Nagasawa and J. Yokoyama,
Phys. Rev. {\bf L77}, 2166 (1996); H.K. Lo, Phys. Rev. {\bf D51}, 7152
(1995).


\bibitem{field} T. Vachaspati and G.B. Field, Phys. Rev. {\bf L73}, 
373 (1994).

\bibitem{barriola} M. Barriola, Phys. Rev. {\bf D51}, R300 (1995). 

\bibitem{super} R.H. Brandenberger and A. Riotto, Phys. Lett.{\bf
    B445}, 323, (1999);  T. Matsuda, Phys. Rev.{\bf D64}, 083512, (2001); 
 L. Masperi, Int. J. Mod. Phys.{\bf A14}, 3581, (1999). 

 
\bibitem{branden} R.H. Brandenberger, A.C. Davis and  M.J. Rees, 
Phys. Lett. {\bf B349} 329, (1995). 

 
\bibitem{jedamzik} K. Jedamzik and G.M. Fuller, Astrophys. J.
{\bf 423}, 33 (1994).

\bibitem{mutc} M. A. Halasz, A. D. Jackson, R. E. Shrock,
  M. A. Stephanov and J. J. M. Verbaarschot, Phys. Rev. {\bf
    D58}, 096007 (1998).   

\bibitem{eta} H. Kurki-Suonio, Phys. Rev. {\bf D37}, 2104 (1988)  

\bibitem{cleyman} J. Cleymans, R.V. Gavai and E. Suhonen,
  Phys. Rept. {\bf130}, 217, (1986)   

\bibitem{lattice} B. Beinlich, F. Karsch, and A. Peikert,
Phys. Lett.  {\bf B390}, 268 (1997).

\bibitem{gavai} B. Banerjee and R. V. Gavai, Phys. Lett. {\bf B293},
  157 (1992).  

\bibitem{kpst} L. P. Csernai and J. I. Kapusta, Phys. Rev. {\bf D46},
  1379 (1992). 

\bibitem{kampfer} B. Kampfer, Annalen Phys. {\bf 9}, 605 (2000) 
 
\bibitem{hogan} A. Heckler and C. J. Hogan, Phys. Rev. {\bf D47},
  4256 (1993)

\bibitem{brn} K. Sumiyoshi, T. Kajino, C. R. Alcock and G. J. Mathews, 
  Phys. Rev. {\bf D42}, 3963 (1990) 

 
\bibitem{stability} M. James, L. Perivolaropoulos and T. Vachaspati. 
Phys. Rev. {\bf D46}, 5232 (1992); Nucl. Phys. {\bf B395}, 534 (1993);
M. Goodband and M. Hindmarsh, hep-ph/9505357; A.Achucarro, R. Gregory,
J.A. Harvey and K. Kuijken, Phys. Rev. {\bf L72}, 3646 (1994);
J. Garriga and X. Montes, Phys. Rev. {\bf L75}, 2268 (1995);
M.Nagasawa and R.Brandenberger hep-ph/9904261;  J.Urrestilla,
A.Achucarro, J.Borrill and A.R. Liddle, hep-ph/0106282.

\bibitem{dw} S.A. Abel and P.L. White, Phys. Rev. {\bf D52},
  4371, (1995); H. Lew and A. Riotto, Phys. Lett. {\bf B309}, 258, (1993).  


\bibitem{bh} Y. Nagatani, Phys. Rev.{\bf D59}, 41301,(1999); see also
    hep-ph/0104160; R. Rangarajan, S. Sengupta and A. M. Srivastava,
    Astropart. Phys. {\bf 17}, 167, (2002).  

\bibitem{fodor} Z. Fodor and S. D. Katz, JHEP {\bf0203}, 014, (2002); 
 Z. Fodor, S. D. Katz and K. K. Szabo, hep-lat/0208078. 






 

\end{thebibliography}
\end{document}